\documentstyle[11pt,newpasp,twoside,epsf]{article}
\begin{document}

\title{High-Mass Stars as Early Signpost of Cluster Formation}
\author{Achim R. Tieftrunk \,\&\, S. Thorwirth}
\affil{KOSMA, I. PI der Universit\"at zu K\"oln, Z\"ulpicher Str. 77, 
50937 K\"oln}
\author{S. T. Megeath}
\affil{Harvard-Smithsonian Center for Astrophysics, 60 Garden Street,
  Cambridge, MA 02138}

\begin{abstract}
  
  The connection between high-mass stars and young stellar clusters has been
  well established by near-IR observations showing young massive stars in
  H~{\sc ii} regions surrounded by clusters of lower mass stars.
  Millimeter-wave observations show that these clusters form in $\approx
  1000$~M$_\odot$ dense cores.  Interestingly, the mm-wave observations of
  very active cluster-forming regions also reveal the presence of seemingly
  {\it quiescent dense massive cores}, which contain few signs of ongoing
  star formation. In the W3~Main region, next to the cluster-forming cores
  W3~SMS1 \& 2, we find a $\approx 1000$~M$_\odot$ mass dense core which
  contains no mid-IR or cm-continuum sources.  Near-IR imaging of this core
  has failed to detect any embedded sources.  Yet, we suggest that this
  region is at an early stage of cluster formation. This is given support by
  comparative observations of a similar twin core system in the NGC 6334
  GMC. Here, north of the highly active star-forming core NGC 6334 I, we
  find another quiescent massive $\approx 1000$~M$_\odot$ core without
  mid-IR sources, H~{\sc ii} regions, or a detectable near-IR cluster.
  Motivated by the presence of CH$_3$OH and H$_2$O masers in this IR-quiet
  core, we have searched for and detected a bipolar outflow driven by a
  young massive star deeply embedded in the core.  The presence of an
  embedded high-mass star strengthens our arguments that IR-quiet cores are
  at an {\it early stage of cluster formation}. We propose that {\it masers
    and outflows} from very deeply embedded (and consequently IR-quiet)
  high-mass stars are among the {\it first incipient signposts} of cluster
  formation in massive cores.  Accordingly, we have undertaken a SEST survey
  of 46 CH$_3$OH maser clusters from the survey of Walsh et~al.\ (1998)
  which are not coincident with IRAS sources or H~{\sc ii} regions.  From
  this sample, we detected outflows in transitions of SiO toward 15 of the
  maser clusters.  These sources will now be observed at 1200~$\mu$m with
  SIMBA, the 37-multichannel bolometer at SEST.

\end{abstract}

\section{Introduction}

High-mass stars, although rare, are a major contributor of energy to the
ambient ISM, be it during their birth in the form of jets and outflows,
during their UV-radiating lifetime, or during their spectacular death as
supernovae. They are commonly detected in rich clusters of stars and to date
an example of isolated high-mass star formation has not been observed.
Thus, if one wants to understand the process of high-mass star formation,
one must understand the physical processes which lead to the formation of
entire clusters of stars.  Furthermore, the dense clusters may influence the
formation of massive stars, and may even be a required precursor for
high-mass star formation.  Is the formation of high-mass stars a natural
implication of the stellar density and the possible interaction and
coalescence of low-mass stars in large clusters?  Do low-mass stars
therefore need to form before a high-mass star can be fused?  Or is it that
high-mass stars simply form independently in the same dense environment as
low-mass stars through the same processes of mass accretion?

In this contribution, we present observations which show the presence
of {\it massive reservoirs of seemingly quiescent dense gas} in close
proximity to highly active cluster-forming cores.  We argue that these
reservoirs are at an early stage of star formation, where the usual
signposts of massive stars and cluster formation -- strong mid-IR
emission, bright radio continuum sources, and dense clusters of stars
in 2~$\mu$m imaging -- are absent.  Instead, the only signatures of
star formation comes from deeply embedded high-mass stars, namely
masers, powerful outflows, and luminous submm sources.  We propose
that {\it some of the youngest regions of cluster formation may be
found by using these early signposts of high-mass star formation}, and
describe an ongoing search for such regions.
  
\section{High-mass Star Formation in Clusters}

From observations of 100 open clusters, Battinelli \& Capuzzo-Dolcetta
(1991) concluded that the star formation rate in clusters in our galaxy is
225~M$_\odot$Myr$^{-1}$ kpc$^{-2}$, whereas the SFR near the sun is
3000--5000~M$_\odot$Myr$^{-1}$kpc$^{-2}$. From observations of Perseus,
L1641, 50\% of L1630, and Mon R2 as part of the Two Micron All Sky Survey,
Carpenter (2000) deduces that the largest fraction of stars must form in
clusters. In fact, 70\% of the young stars identified in Carpenter's
analysis are in clusters with more than 100 members.  This then leads one to
the conclusion that the 10\% still seen as open clusters are simply those
that remain bound long after their parental molecular clouds are dissipated.
The dispersal of the molecular gas in clusters is very much a function of
their most massive members.  High-mass stars ($>5$ M$_\odot$, corresponding
to spectral type B6 or earlier) are always found to be associated with rich
clusters with stellar densities of more than 1000 stars per pc$^{3}$.  These
clusters are thus affected by the dispersive jets, the intense UV fields
and, if the cluster remains bound after the dispersal of the molecular
cloud, the ensuing supernovae of the massive stars.

From observation of Herbig Ae/Be stars, Testi, Palla, \& Natta (1999)
conclude that the lack of clusters around stars later than B7--B8 is an
imprint of the stellar formation mode for high-mass stars.  The process of
high-mass star formation within dense clusters through spherical accretion
has been a long-standing problem. At around 8--10~M$_\odot$ the radiation
pressure of the forming star begins to inhibit further infall of the
material accreting (e.g.~Yorke \& Kruegel 1977, Kahn 1974, Larson \&
Starrfield 1971). Models including rotating protostellar grainy disks
(e.g.~Jijina \& Adams 1996, Wolfire \& Cassinelli 1987), magnetic fields
(Balsara et~al.\ 2001, Nakano 1998), and accretion rates growing with time
(Behrend \& Maeder 2001) can certainly push this limit very much higher, but
the theories about the physical backgrounds for the accretion process are
still open for much discussion.  At stellar densities of around
10$^7$pc$^{-3}$ accretion induced collision may begin to have a significant
effect on the formation of high-mass stars up to the point where fusion
of lower mass stars may actually form the more massive members (e.g.~Bonnell
et~al.\ 1998, Stahler et~al.\ 1998). These two scenarios for forming massive
stars in clusters -- accretion and star coalescence -- are the focus of the
current debate.

\section{Early Signposts of High-Mass Stars}

High-mass stars are typically formed in warm ($\approx 30$~K), dense
(n(H$_2$) $\approx 10^5$ cm$^{-3}$), compact ($\approx 50,000$~AU) and
massive ($\approx 1000$~M$_\odot$) dust cores, easily traced in CS,
CH$_3$CN, HC$_3$N, or NH$_3$ and the submm-continuum with typical FIR
luminosities of $10^3-10^5$~L$_\odot$. These cores are usually associated
with H$_2$O, CH$_3$OH, H$_2$CO, NH$_3$, and/or OH masers and often, at a
later stage when the most massive members are emerging from their natal
cocoons, hyper- or ultracompact H~{\sc ii} regions.  Often one can find
shock-excited H$_2$ emission knots at 2.1~$\mu$m and/or high-velocity
bipolar outflows from broad wings in SiO or $^{13}$CO molecular emission
lines.  These molecular outflows typically have masses of around 150
(10--5000)~M$_\odot$ and outflow rates of $3\times 10^{-5} - 3\times
10^{-2}$~M$_\odot$yr$^{-1}$ (Garay \& Lizano 1999). Outflows and certain
types of masers appear very early on in the formation process, long before
luminous IR sources or H~{\sc ii} regions can be detected.  Sheperd \&
Churchwell (1996) have shown that the outflow properties seem to be
correlated with the luminosity of the embedded sources driving the outflows.
Although new observation by Henning et al.\ (2000) are in support of this, a
recent analysis of a number of massive outflows seems to indicate that this
empirical relationship breaks down with increasingly higher mass (cf.\ 
contribution by Beuther {et~al.}, this issue).

\subsection{W3~SMS3}

W3 Main is a well-studied and very active region of high-mass star formation
in the W3 GMC at a distance of 2.3 kpc. The W3 Main region is 10 pc north of
the W3(OH) region and consists of two main molecular cores: W3~East \& West.
Towards W3~East we find a dense stellar cluster associated with the luminous
infrared source IRS~5 embedded in the compact molecular core W3~SMS1
(Tieftrunk et~al.\ 1998).  Towards this region Megeath et~al.\ (1996)
detected the highest density of stars: a cluster of $\approx 80-300$
low-mass stars, corresponding to a stellar density of a few $10^3$
pc$^{-3}$.  The bright source IRS~5 itself is associated with a massive
bipolar outflow (Hasegawa et~al.\ 1994) and a cluster of H$_2$O masers where
Claussen et~al.\ (1994) have detected a number of hypercompact H~{\sc ii}
regions with diameters less than 1000 AU.  Towards IRS~4, embedded in the
compact molecular core W3~SMS2, Tieftrunk et~al.\ (1997) found another
similar hypercompact H~{\sc ii} region.  The presence of hypercompact
sources indicates that the massive stars are just beginning to ionize their
surrounding cores (Tieftrunk et~al.\ 1997).  Just to the south of IRS~4, is
the molecular core W3~West/SMS3 (Ladd et~al.\ 1993).  This dense core is
seemingly quiescent: it is devoid of mid-IR sources and radio continuum
sources, and 2~$\mu$m imaging has failed to detect a cluster in SMS3
(Tieftrunk et al.\ 1998).  Since SMS3 has a molecular mass and density
higher than the surrounding cluster-forming cores, Tieftrunk, Gaume \&
Wilson (1998) argued that the SMS3 core may be at an early stage of cluster
formation.

\begin{figure}[ttt]
\vspace{5cm}
\caption{Contour plot of a 6 cm continuum VLA image superimposed on the
  C$^{18}$O emission from Tieftrunk et~al.\ (1997).
  Contours are $(2-20)\times 2, 24, 30, 40, 60, 80,$ and 100 mJy/beam.
  Continuum components and 450~$\mu$m sources indicated by crosses
  are labeled.}
\end{figure}

\begin{figure}[bbb]
\plotfiddle{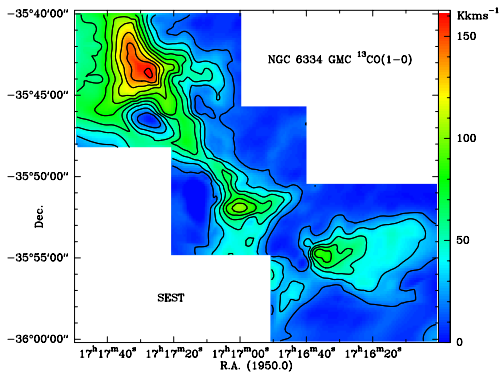}{4cm}{-90}{30}{30}{-90}{155}
\caption{$^{13}$CO 1-0 emission mapped with the SEST towards
  the giant molecular cloud NGC 6334. The twin cores NGC 6334 I \& I(N)
  are found in the elongated structure towards the north-east.}
\end{figure}

\subsection{NGC 6334 I(N)}

Towards the north-eastern edge of the giant molecular cloud NGC 6334
(Fig.~2), at a distance of 1.7 kpc (Neckel 1978), another massive twin core
system can be found: NGC 6334 I and I(N) (Fig.~3). As W3~SMS1 \& 3, these
two cores present a unique duet for studying the evolution of stellar
clusters and the interstellar chemistry in young star-forming cores.

\begin{figure}
  \plotfiddle{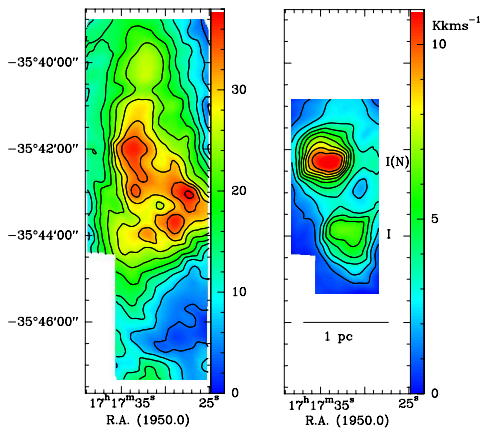}{5cm}{-90}{35}{35}{-110}{165}
\caption{C$^{18}$O (2-1) [left] and HC$_3$N (15-14) [right] emission 
  mapped with SEST toward the twin core system NGC 6334 I \& I(N).}
\end{figure}

NGC 6334 I, with a total mass of $\approx 1000$~M$_\odot$ and a dust
temperature of $\approx 100$ K, is a dense and compact molecular hot core
associated with an ultracompact cometary H~{\sc ii} as well as a strong IR
continuum region and a group of CH$_3$OH, H$_2$O, OH, and NH$_3$ masers.  It
is considered to be the youngest of the star-forming regions in the NGC 6334
GMC, exhibiting the richest chemistry of all sources. In fact, our SEST data
shows that this source is comparable to Orion-KL and Sgr B2 in the richness
of its spectrum.  However, in comparison to these sources where the
preponderance of line blending is a principal problem for assigning emission
lines or detecting new transitions, line widths in NGC 6334~I are 2--4 times
more narrow allowing for a significantly higher detection rate of weaker
features (Fig.~4). In fact, we detected a plethora of line features down to
a noise level of 0.01 K and expect that {\it Sgr B2 and Orion-KL will be
  replaced by NGC 6334 I as the new standard source in searches for new
  interstellar molecules and transitions} (Tieftrunk \& Thorwirth, in
prep.).

\begin{figure}[hhh]
\plotfiddle{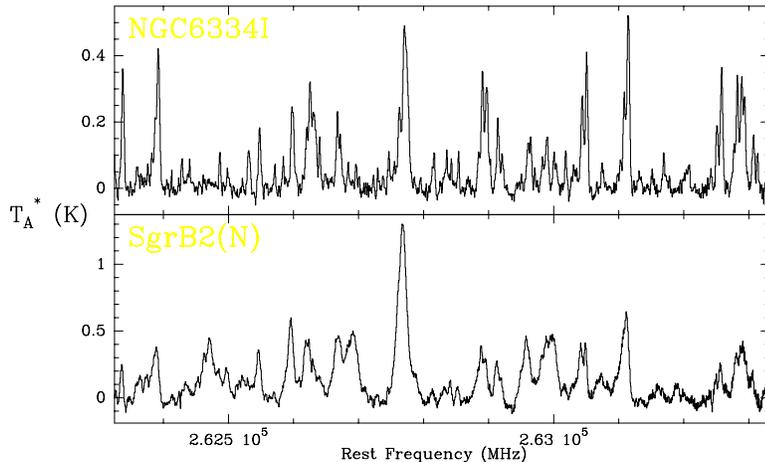}{5cm}{-90}{40}{40}{-160}{200}
\caption{Sample spectrum of 1 GHz bandwidth around 263~GHz towards
  SgR B2(N) [bottom] and NGC 6334 I [top] taken from our survey
  covering approximately 140 GHz between 78 GHz and 268 GHz.}
\end{figure}

Only 1.5$^{\prime}$ north (1~pc at 1.7~kpc) of NGC 6334 I a second core of
roughly equal size, comparable mass, and a molecular hydrogen density of
$\approx 10^6$ cm$^{-3}$ (Kuiper et al.\ 1995) is apparent (Sandell 2000).
Designated NGC 6334 I/(North) [hereafter I(N)] it is notable for the lack of
mid-IR emission: to date, no thermal infrared emission has been detected
from this region at wavelengths $< 100 \mu$m (Gezari 1982).  It does not
contain an H~{\sc ii} region.  However, H$_2$O and CH$_3$OH (class I and II)
masers have been detected in the core.  Very recently, Megeath \& Tieftrunk
(1999) presented evidence from SEST and ESO-2.2m data for multiple outflows
originating in I(N) (Fig.  5).  Using the SiO and $^{13}$CO emission in the
line wings towards I(N), we deduced an outflow mass of 7~M$_\odot$,
transporting $\approx 10^{-3}$~M$_\odot$ yr$^{-1}$.  From the well-known
relationship of Shepherd \& Churchwell (1996) we thus estimated a luminosity
of $10^3$~L$_\odot$ and a mass of 10~M$_\odot$ for the driving source.
Thus, the I(N) core already contains at least one high-mass (proto)star and
is possibly on the verge of cluster formation (cf. also Sandell 2000).  From
these observations we conclude that the {\it first detectable evidence for
  cluster formation are the signposts of outflows and masers of the
  associated high-mass stars} long before a dense 2~$\mu$m cluster itself
appears. Thus the formation of intermediate to high-mass stars can at least
precede the {\it appearance} of the associated cluster of low-mass stars.
This, however, does not imply that low-mass stars would actually form later
as well; they may be too deeply embedded to become visible to the observer
at an early stage of the cluster formation.

\begin{figure}
\plotfiddle{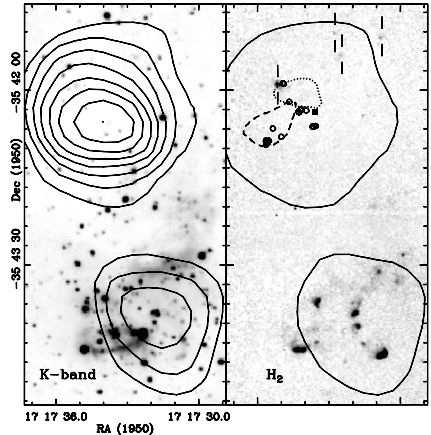}{5cm}{-90}{40}{40}{-100}{200}
\caption{{\sc left}: K-band image with contours of HC$_3$N(15-14) emission
  towards NGC 6334 I \& I(N); {\sc right}: continuum subtracted
  H$_2$($\nu=1-0$)S(1) emission (marked with vertical lines) and 
  CH$_3$OH maser positions (dots) towards I(N) with contours of
  blue (dashed) and red (dotted) SiO(5-4) line wings and the lowest
  HC$_3$N(15-14) contour.}
\end{figure}

\section{IRAS-quiet maser clusters}

Surveys of the earliest stages of massive star formation have
concentrated on UC H~{\sc ii} regions (e.g.~Brand et~al.\ 1984, Wood
\& Churchwell 1989), some associated with IRAS sources (e.g.~Bronfman
et~al.\ 1996, Osterloh et~al.\ 1997), and more recently, on luminous
IRAS sources which are not associated with H~{\sc ii} regions
(e.g.~Zhang et~al.\ 2001, Molinari et~al.\ 2000, Sridharan et~al.\
1999).  However, as demonstrated by NGC 6334 I(N) and W3~SMS3, the
earliest phases of massive star formation are probably both radio and
infrared-quiet -- analogous to the Class 0 phase of low-mass star
formation -- and searches using the IRAS database may miss some of the
youngest, most interesting sources.  Due to the lack of unbiased
submillimeter surveys of the galaxy, only a few truly deeply embedded
massive protostars or Class 0 objects are known.

From the ATCA survey by Walsh et~al.\ (1998) we had selected 46 maser
clusters that were radio- and IR-quiet, i.e. not associated with UC H~{\sc
  ii} regions and more than 1 pc (1\arcmin --14\arcmin) away from the
nearest IRAS point source. With the SEST we detected broad SiO emission line
wings towards 15 of these, indicative of outflows. It is clear that these
CH$_3$OH maser clusters with SiO outflows must have a driving source.  These
sources, however, are inconspicuous in the mid-IR and thus must be deeply
embedded. NGC 6334 I(N), which has a comparable flux density to NGC 6334 I
at 400~$\mu$m, is much brighter at 1 mm (Gezari 1982). Thus, these deeply
embedded objects appear to be the prominent sources when observed at 1~mm in
a system of star-forming cores.  With SIMBA, the 37-multichannel bolometer
of SEST, we will now observe the 1200~$\mu$m flux of the maser sources.
Since dust emission at (sub)mm wavelengths is generally optically thin, it
provides a much better contrast between high and low density gas than most
molecular lines. From this sample of regions, we expect to find not only the
youngest examples of high-mass star formation, but some of the youngest
regions of cluster formation.  With such a sample we can better study the
relationship between high- and low-mass stars, as well as construct a better
picture of the physical and chemical environment of molecular clouds at the
onset of cluster formation.


\begin{references}
\reference
Balsara, D. S., Ward-Thompson, D., \& Crutcher, R. M. 2001, \mnras, 327, 715
\reference
Battinelli, P. \& Capuzzo-Dolcetta, R. 1991, \mnras, 249, 76
\reference
Behrend, R. \& Maeder, A. 2001, \aap, 373, 190
\reference
Bonnell, I. A., Bate, M. R., \& Zinnecker, H. 1998, \mnras, 298, 93
\reference
Brand, J., van der Bij, M. D. P., de Vries, C. P., Leene, A., Habing, H. J.,
Israel, F. P., de Graauw, T., van de Stadt, H. \& Wouterloot, J. G. A.
1984, \aap, 139, 181
\reference
Bronfman, L., Nyman, L.-A., \& May, J. 1996, \aaps, 115, 81
\reference
Carpenter, J. M. 2000, \aj, 120, 3139
\reference
Claussen, M. J., Gaume, R. A., Johnston, K. J., \& Wilson, T. L. 1994, 
\apj, 424, L41
\reference
Garay, G. \& Lizano, S. 1999, \pasp, 111, 1049
\reference
Gezari, D. Y. 1982, \apj, 259, L29
\reference
Hasegawa, T. I., Mitchell, G. F., Matthews, H. E., \& Tacconi, L. 1994, 
\apj, 426, 215
\reference
Henning, Th., Schreyer, K., Launhardt, R., \& Burkert, A. 2000, \aap, 353,
211
\reference
Jijina, J. \& Adams, F. C. 1996, \apj, 462, 874
\reference
Kahn, F. D. 1974, \aap, 37, 149
\reference
Kuiper, T. B. H., Peters, W. L., III, Foster, J. R., Gardner, F. F., \&
Whiteoak, J. B. 1995, \apj, 446, 692
\reference
Ladd, E. F., Deane, J. R., Sanders, D. B., Wynn-Williams, C. G. 1993, 
\apj, 419, L186
\reference
Larson, R. B. \& Starrfield, S. 1971, \aap, 13, L190
\reference
Megeath, S. T. \& Tieftrunk, A. R. 1999, \apj, 526, L113
\reference
Megeath, S. T., Herter, T., Beichman, C., Gautier, N., Hester, J. J.,
Rayner, J., \& Shupe, D. 1996, \aap, 307, 775
\reference
Molinari, S., Brand, J., Cesaroni, R., \& Palla, F. 2000, \aap, 355, 617
\reference
Nakano, T. 1998, \apj, 494, 587
\reference
Neckel, T. 1978, \aap, 69, 51
\reference
Osterloh, M., Henning, Th., \& Launhardt, R. 1997, \apjs, 110, 71
\reference
Sandell, G. 2000, \aap, 358, 242
\reference
Shepherd, D. S. \& Churchwell, E. 1996, \apj, 472, 225
\reference
Stahler, S. W., Palla, F., \& Ho, P. T. 1998, in: Protstars and Planets IV,
eds. Mannings, V. {et~al.}\, Univ. of Arizona Press, Tucson, 397
\reference
Sridharan, T. K., Menten, K. M., Wyrowski, F., \& Schilke, P. 1999, in:
Proceedings of Star Formation 1999, eds. Nakamoto, T., Nobeyama Radio
Observatory, 183
\reference
Testi, L., Palla, F., \& Natta, A. 1999, \aap, 342, 515
\reference
Tieftrunk, A. R., Gaume, R. A., Claussen, M. J., Wilson, T. L., \& Johnston,
K. J. 1997, \aap, 318, 931
\reference
Tieftrunk, A. R., Gaume, R. A., \& Wilson, T. L. 1998, \aap, 340, 232
\reference
Tieftrunk, A. R., Megeath, S. T., Wilson, T. L., \& Rayner, J. T. 1998,
\aap, 336, 991
\reference
Walsh, A. J., Burton, M. G., Hyland, A. R., \& Robinson, G. 1998, 
\mnras, 301, 640
\reference
Wolfire, M. G. \& Cassinelli, J. P. 1987, \apj, 319, 85
\reference
Wood, D. O. S. \& Churchwell, E. 1989, \apjs, 69, 831
\reference
Yorke, H. W. \& Kruegel, E. 1977, \aap, 54, 183
\reference
Zhang, Q., Hunter, T. R., Brand, J., Sridharan, T. K., Molinari, S.,
Kramer, M. A., \& Cesaroni, R. 2001, \apj, 552, 167
\end{references}
\end{document}